# Some ambiguities with point split regularization and its impact on a proof of the spatial quantum inequality.

Dan Solomon
Rauland-Borg Corporation
Mount Prospect, IL
Email: dan.solomon@rauland.com
March 9, 2011

**Abstract**.

In classical physics the energy density of a field is always positive. However this does not hold true for quantum physics where the energy density of a field can be locally negative. There are limits on the weighted average of this negative energy density called the quantum inequalities. Recently this author has provided a number of examples which show that the quantum inequalities are not valid. In this paper we will examine a previously published proof of the spatial quantum inequality for a zero mass scalar field in 1-1 dimensional space-time. It will be shown that there is a possible problem with this proof due to an ambiguity associated with point split regularization and the definition of the Hadamard form for the two point function.

## 1. Introduction.

The energy density of a classical field is always positive. However this is not necessarily the case in quantum field theory. In quantum field theory it has been shown that the energy density can be negative [1]. It has been suggested that the existence of negative energy could result in a violation of the second law of thermodynamics [2] and could give rise to unusual effects such as traversable wormholes [3]. However are a number of papers that claim there are limits on this effect [4-8] that would prevent these "exotic" phenomenon from occurring. These limits are referred to as the quantum inequalities. The quantum inequalities provide a lower boundary on the weighted average of the energy density over some region of space and time. They apply to free field systems, that is, systems where all external potentials are zero.

   E. E. Flanagen [5] examined quantum inequalities in 1-1 dimensional space-time for a massless scalar field. According to [5] the spatial quantum inequality is given by,

$$\int_{-\infty}^{+\infty} T_{00}(x,t)\rho(x)dx \geq -\frac{1}{24\pi}\int_{-\infty}^{+\infty} dx\frac{\rho'(x)^2}{\rho(x)} \qquad (1.1)$$

where $T_{00}(x,t)$ is the energy density and $\rho(x)$ is a strictly positive weighting function. There is a similar relationship for the temporal quantum inequality except the integrations are over time instead of space.

Recently a number of papers have been written by this author that claim to demonstrate counterexamples to the quantum inequalities (D. Solomon [9-12]). This implies that there is something wrong with the proofs that claim to support the quantum inequalities. In this paper we will focus on the spatial quantum inequality as discussed by Flanagan [5]. It will be shown that there is a possible problem with the proof presented in [5] due to the way the limits are taken in the Hadamard form of the two point function.

Consider scalar field theory in 1-1D space-time. Let $x$ be the space coordinate and $t$ the time coordinate. In this case the Hadamard form of the two point function $G(x,t;x',t')$ can be expressed as [13, 14],

$$G(x,t;x',t') = U(x,t;x',t')\ln\sigma_1 + W(x,t;x',t') \qquad (1.2)$$

where,

$$\sigma_1 = \sigma + 2i(t-t')\tau + \tau^2; \quad \sigma = (x-x')^2 - (t-t')^2 \qquad (1.3)$$

The quantity $\tau \to 0$ is assumed to approach zero and is required in order to prevent $\ln\sigma_1$ from being undefined when $(x-x')^2 - (t-t')^2 = 0$. (See discussion in Section 4.6 of Wald [14]).

The two point function is used to determine the energy density using the method of point split regularization. In this case, at the end of the calculations, we take, $x' \to x$ and $t' \to t$. Note that this means we will have three quantities that are approaching zero. These are $(x-x')$, $(t-t')$, and $\tau$. It will be shown that final results are dependent on the order in which we let these quantities go to zero. This will impact on Flanagan's proof of the spatial quantum inequality.

In the following discussion we will do some fairly straightforward calculations involving the kinetic energy density using point split regularization and show how the results are dependent on the way in which the quantities $(x-x')$, $(t-t')$, and $\tau$ approach zero. Then we will argue that the proper way to take these terms to zero is to allow $(x-x')$ and $(t-t')$ to approach zero much faster than $\tau$. It will be shown that if this is done Flanagan's proof of the spatial quantum inequality no longer holds.

## 2. Calculation involving a massive scalar field.

For the first example consider a massive scalar field of in 1-1 dimensional space-time in the presence of the time independent scalar potential $\lambda V(x)$. In this case the field operator $\hat{\varphi}(x,t)$ satisfies the Klein-Gordon equation,

$$\frac{\partial^2 \hat{\varphi}_\lambda}{\partial t^2} - \frac{\partial^2 \hat{\varphi}_\lambda}{\partial x^2} + m^2 + \lambda V(x)\hat{\varphi}_\lambda = 0 \tag{2.1}$$

where $m$ is the mass and $\lambda$ is a non-negative parameter that can be set to zero to turn off the potential. The field operator is,

$$\hat{\varphi}_\lambda(x,t) = \sum_n \left( \hat{a}_{\lambda,n} f_{\lambda,n}(x,t) + \hat{a}^*_{\lambda,n} f^*_{\lambda,n}(x,t) \right) \tag{2.2}$$

where $\hat{a}_{\lambda,n}$ and $\hat{a}^*_{\lambda,n}$ are the destruction and creation operators, respectively. They obey the commutation relationships $\left[ \hat{a}_{\lambda,n}, \hat{a}^*_{\lambda,n'} \right] = \delta_{n'n}$ with all other commutations equal to zero. The mode solutions $f_{\lambda,n}(x,t)$ satisfy Eq. (2.1) and the usual normalization condition (See section 2.1 of [15]).

Define the normalized state vector $|0_\lambda\rangle$ by the relationships,

$$\hat{a}_{n,\lambda} |0_\lambda\rangle = \langle 0_\lambda | \hat{a}^\dagger_{n,\lambda} = 0 \text{ and } \langle 0_\lambda | 0_\lambda \rangle = 1 \tag{2.3}$$

$|0_\lambda\rangle$ may be thought of as the vacuum state in the presence of the scalar potential $\lambda V(x)$.

Next we want to calculate the kinetic energy density for the state $|0_\lambda\rangle$. The kinetic energy operator is defined by,

$$\hat{T}_{00}(x,t;[\hat{\varphi}_\lambda]) = \frac{1}{2} \left( \frac{\partial \hat{\varphi}_\lambda}{\partial t} \frac{\partial \hat{\varphi}_\lambda}{\partial t} + \frac{\partial \hat{\varphi}_\lambda}{\partial x} \frac{\partial \hat{\varphi}_\lambda}{\partial x} + m^2 \hat{\varphi}_\lambda \hat{\varphi}_\lambda \right) \tag{2.4}$$

The kinetic energy density expectation value of the state $|0_\lambda\rangle$ is, then, given by,

$$T_{00,\lambda}(x,t) = \langle 0_\lambda | \hat{T}_{00}(x,t;[\hat{\varphi}_\lambda]) | 0_\lambda \rangle \tag{2.5}$$

Now there is a problem with this evaluating this expression. It can easily be shown to be infinite. However we are not really interested in absolute magnitude of the kinetic energy density but in the difference between this kinetic energy density and the kinetic energy density of the unperturbed vacuum state. Therefore we define the regularized kinetic energy density by the expression,

$$T_{00R}(x,t) = \langle 0_\lambda | \hat{T}_{00}(x,t;[\hat{\varphi}_\lambda]) | 0_\lambda \rangle - \langle 0_0 | \hat{T}_{00}(x,t;[\hat{\varphi}_0]) | 0_0 \rangle \tag{2.6}$$

where $\hat{\varphi}_0$ is "free field" solution of the field operator for the case where the scalar potential is zero. The quantities $\hat{\varphi}_0$ and $|0_0\rangle$ are defined by setting $\lambda = 0$ in Eqs. (2.1) through (2.3).

There is still a problem with evaluating this expression due to the fact that we are subtracting one infinity from another. We will ignore this problem for the moment. Next, define the momentum density operator as

$$\hat{T}_{01}(x,t;[\hat{\varphi}_\lambda]) = \frac{1}{2}\left( \frac{\partial \hat{\varphi}_\lambda}{\partial t} \frac{\partial \hat{\varphi}_\lambda^*}{\partial x} + \frac{\partial \hat{\varphi}_\lambda}{\partial x} \frac{\partial \hat{\varphi}_\lambda^*}{\partial t} \right) \tag{2.7}$$

Following the procedure leading up to (2.6) we define the renormalized momentum density expectation value of the state $|0_\lambda\rangle$ as,

$$T_{01R}(x,t) = \langle 0_\lambda | \hat{T}_{01}(x,t;[\hat{\varphi}_\lambda]) | 0_\lambda \rangle - \langle 0_0 | \hat{T}_{01}(x,t;[\hat{\varphi}_0]) | 0_0 \rangle \tag{2.8}$$

Now evaluate the expression $\partial T_{01R}(x,t)/\partial t$. This will be done using the equation of motion (2.1) in (2.8) along with the fact that, for the free field, $\langle 0_0 | \hat{T}_{01}(x,t;[\hat{\varphi}_0]) | 0_0 \rangle$ and $\langle 0_0 | \hat{\varphi}_0^*(x,t) \hat{\varphi}_0(x,t) | 0_0 \rangle$ are space and time independent. We obtain,

$$\frac{\partial T_{01R}(x,t)}{\partial t} = \frac{\partial T_{00R}(x,t)}{\partial x} - \left( m^2 + \frac{\lambda V(x)}{2} \right) \frac{\partial F_R(x,t)}{\partial x} \tag{2.9}$$

where,

$$F_R(x,t) = \langle 0_\lambda | \hat{\varphi}_\lambda^*(x,t) \hat{\varphi}_\lambda(x,t) | 0_\lambda \rangle - \langle 0_0 | \hat{\varphi}_0^*(x,t) \hat{\varphi}_0(x,t) | 0_0 \rangle \tag{2.10}$$

As discussed above there is a potential problem with this calculation in that some of the quantities, such as $\langle 0_\lambda | \hat{T}_{00}(x,t;[\hat{\varphi}_\lambda]) | 0_\lambda \rangle$ and $\langle 0_\lambda | \hat{\varphi}_\lambda^*(x,t) \hat{\varphi}_\lambda(x,t) | 0_\lambda \rangle$ are

divergent and, therefore, not well defined. In order to resolve this problem we will redefine the kinetic energy density and momentum density operators using point splitting and then repeat the above calculations using the point split operators. First, in order to simplify the notation, define the following ordered pairs, $y = (y_0, y_1)$ and $y' = (y'_0, y'_1)$ where,

$$y_0 = t + \varepsilon_0/2; \quad y'_0 = t - \varepsilon_0/2; \quad y_1 = x + \varepsilon_1/2; \quad y'_1 = x - \varepsilon_1/2 \tag{2.11}$$

The point split kinetic energy density operator is then defined by,

$$\hat{T}_{00}(y, y'; [\hat{\varphi}_\lambda]) = \frac{1}{2}\left(\hat{t}_{00}(y, y'; [\hat{\varphi}_\lambda]) + \hat{t}_{00}(y', y; [\hat{\varphi}_\lambda])\right) \tag{2.12}$$

where,

$$\hat{t}_{00}(y, y'; [\hat{\varphi}_\lambda]) = \frac{1}{2}\left(\frac{\partial \hat{\varphi}_\lambda(y)}{\partial y_0}\frac{\partial \hat{\varphi}^*_\lambda(y')}{\partial y'_0} + \frac{\partial \hat{\varphi}_\lambda(y)}{\partial y_1}\frac{\partial \hat{\varphi}^*_\lambda(y')}{\partial y'_1} + m^2 \hat{\varphi}_\lambda(y)\hat{\varphi}^*_\lambda(y')\right) \tag{2.13}$$

In this case the kinetic energy density expectation value is given by,

$$T_{00R}(x,t;\varepsilon) = \langle 0_\lambda | \hat{T}_{00}(y, y'; [\hat{\varphi}_\lambda]) | 0_\lambda \rangle - \langle 0_0 | \hat{T}_{00}(y, y'; [\hat{\varphi}_0]) | 0_0 \rangle \tag{2.14}$$

where $T_{00R}(x,t;\varepsilon)$ is potentially dependent on the pair $\varepsilon = (\varepsilon_0, \varepsilon_1)$. If point split regularization works then the terms containing $\varepsilon$ should drop out in the limit that $\varepsilon = (\varepsilon_0, \varepsilon_1) \to 0$.

Eq. (2.14) can be more conveniently expressed in terms of the two point function which is defined as,

$$G_\lambda(y; y') = \frac{1}{2}\langle 0_\lambda |(\hat{\varphi}_\lambda(y)\hat{\varphi}_\lambda(y') + \hat{\varphi}_\lambda(y')\hat{\varphi}_\lambda(y))| 0_\lambda \rangle \tag{2.15}$$

The two point function for the free field case is obtained by setting $\lambda = 0$ in the above expression to obtain,

$$G_0(y; y') = \frac{1}{2}\langle 0_0 |(\hat{\varphi}_0(y)\hat{\varphi}_0(y') + \hat{\varphi}_0(y')\hat{\varphi}_0(y))| 0_0 \rangle \tag{2.16}$$

Note that both $G_\lambda(y; y')$ and $G_0(y; y')$ will be well defined expressions that will be dependent on $\varepsilon$. The point split renormalized kinetic energy density is given by,

$$T_{00R}(x,t;\varepsilon) = T_{00,\lambda}(x,t;\varepsilon) - T_{00,0}(x,t;\varepsilon) \tag{2.17}$$

where,

$$T_{00,\lambda}(x,t;\varepsilon) = \frac{1}{2}\left(\frac{\partial}{\partial y_0}\frac{\partial}{\partial y_0'} + \frac{\partial}{\partial y_1}\frac{\partial}{\partial y_1'} + m^2\right) G_\lambda(y,y') \qquad (2.18)$$

and,

$$T_{00,0}(x,t;\varepsilon) = \frac{1}{2}\left(\frac{\partial}{\partial y_0}\frac{\partial}{\partial y_0'} + \frac{\partial}{\partial y_1}\frac{\partial}{\partial y_1'} + m^2\right) G_0(y,y') \qquad (2.19)$$

Similarly the renormalized momentum density is given by,

$$T_{01R}(x,t;\varepsilon) = T_{01,\lambda}(x,t;\varepsilon) - T_{01,0}(x,t;\varepsilon) \qquad (2.20)$$

where,

$$T_{01,\lambda}(x,t;\varepsilon) = \frac{1}{2}\left(\frac{\partial}{\partial y_1}\frac{\partial}{\partial y_0'} + \frac{\partial}{\partial y_0}\frac{\partial}{\partial y_1'}\right) G_\lambda(y,y') \qquad (2.21)$$

and

$$T_{01,0}(x,t;\varepsilon) = \frac{1}{2}\left(\frac{\partial}{\partial y_1}\frac{\partial}{\partial y_0'} + \frac{\partial}{\partial y_0}\frac{\partial}{\partial y_1'}\right) G_0(y,y') \qquad (2.22)$$

Also define,

$$F_R(x,t;\varepsilon) = G_\lambda(y,y') - G_0(y,y') \qquad (2.23)$$

From (2.11) we obtain,

$$\frac{\partial}{\partial t} \to \frac{1}{2}\left(\frac{\partial}{\partial y_0} + \frac{\partial}{\partial y_0'}\right) \quad ; \frac{\partial}{\partial x} \to \frac{1}{2}\left(\frac{\partial}{\partial y_1} + \frac{\partial}{\partial y_1'}\right) \qquad (2.24)$$

Use this in (2.21) to yield,

$$\frac{\partial}{\partial t} T_{01,\lambda}(x,t;\varepsilon) = \frac{1}{4}\left\{\left(\frac{\partial}{\partial y_1}\frac{\partial^2}{\partial y_0'\partial y_0} + \frac{\partial^2}{\partial y_0'\partial y_0}\frac{\partial}{\partial y_1'}\right) + \left(\frac{\partial}{\partial y_1}\frac{\partial^2}{\partial y_0'^2} + \frac{\partial^2}{\partial y_0^2}\frac{\partial}{\partial y_1'}\right)\right\} G_\lambda(y,y') \qquad (2.25)$$

Next use the fact that the two point function $G_\lambda(y,y')$ obeys the equation of motion (2.1) in the above to obtain,

$$\frac{\partial}{\partial t} T_{01,\lambda}(x,t;\varepsilon) = \frac{\partial T_{00,\lambda}(x,t;\varepsilon)}{\partial x} - m^2 \frac{\partial}{\partial x} G_\lambda(y,y') - \frac{\lambda}{4}\left(V(y_1')\frac{\partial}{\partial y_1} + V(y_1)\frac{\partial}{\partial y_1'}\right) G_\lambda(y,y')$$

(2.26)

Set $\lambda = 0$ in the above to obtain,

$$\frac{\partial}{\partial t} T_{01,0}(x,t;\varepsilon) = \frac{\partial T_{00,0}(x,t;\varepsilon)}{\partial x} - m^2 \frac{\partial}{\partial x} G_0(y,y') \qquad (2.27)$$

Therefore,

$$\frac{\partial}{\partial t}T_{01R}(x,t;\varepsilon) = \frac{\partial T_{00R}(x,t;\varepsilon)}{\partial x} - m^2 \frac{\partial}{\partial x}\left(G_\lambda(y,y') - G_0(y,y')\right)$$
$$-\frac{\lambda}{4}\left(V(y'_1)\frac{\partial}{\partial y_1} + V(y_1)\frac{\partial}{\partial y'_1}\right)G_\lambda(y,y') \tag{2.28}$$

Consider the last term. Recall that we are evaluating these expressions in the limit $\varepsilon \to 0$ so that we can write,

$$V(y_1) = V(x + \varepsilon_1/2) = V(x) + \frac{\varepsilon_1}{2}\frac{\partial V(x)}{\partial x} \tag{2.29}$$

and,

$$V(y'_1) = V(x - \varepsilon_1/2) = V(x) - \frac{\varepsilon_1}{2}\frac{\partial V(x)}{\partial x} \tag{2.30}$$

Therefore,

$$\left(V(y'_1)\frac{\partial}{\partial y_1} + V(y_1)\frac{\partial}{\partial y'_1}\right)G_\lambda(y,y') = \left\{\begin{array}{l} V(x)\left(\dfrac{\partial}{\partial y_1} + \dfrac{\partial}{\partial y'_1}\right) \\ +\dfrac{\varepsilon_1}{2}\dfrac{\partial V(x)}{\partial x}\left(\dfrac{\partial}{\partial y_1} - \dfrac{\partial}{\partial y'_1}\right) \end{array}\right\} G_\lambda(y,y') \tag{2.31}$$

Assume that the two point function is of the Hadamard form. In this case,

$$G_\lambda(y,y') = U_\lambda(y,y')\ln\sigma_1 + W_\lambda(y,y') \tag{2.32}$$

where,

$$\sigma_1 = \sigma + 2i(y_0 - y'_0)\tau + \tau^2; \quad \sigma = (y_1 - y'_1)^2 - (y_0 - y'_0)^2 \tag{2.33}$$

with $\tau \to 0$. Also from [13,14],

$$U_\lambda(y,y') = 1 + u_{1,\lambda}(y,y')\sigma + u_{2,\lambda}(y,y')\sigma^2 + \ldots \tag{2.34}$$

and,

$$W_\lambda(y,y') = w_{0,\lambda}(y,y') + w_{1,\lambda}(y,y')\sigma + w_{2,\lambda}(y,y')\sigma^2 + \ldots \tag{2.35}$$

Using the above relationships we obtain,

$$\frac{\varepsilon_1}{2}\left(\frac{\partial}{\partial y_1} - \frac{\partial}{\partial y'_1}\right)G_\lambda(y,y')\bigg|_{\sigma_1 \to 0} = \frac{2\varepsilon_1^2}{\sigma_1} \tag{2.36}$$

where the expression $\sigma_1 \to 0$ implies that the three quantities $\varepsilon_0$, $\varepsilon_1$, and $\tau$ all approach zero.

For the free field case we have $G_0(y, y') = G_0(y - y')$. Therefore,

$$\frac{\partial}{\partial x} G_0(y - y') = \frac{1}{2}\left(\frac{\partial}{\partial y_1} + \frac{\partial}{\partial y_1'}\right) G_0(y - y') = 0 \tag{2.37}$$

From all this we obtain,

$$\frac{\partial}{\partial t} T_{01R}(x,t;\varepsilon) = \frac{\partial T_{00R}(x,t;\varepsilon)}{\partial x} - \left(\frac{\lambda V(x)}{2} + m^2\right)\frac{\partial}{\partial x} F_R(x,t;\varepsilon) - \frac{\lambda}{2}\frac{\partial V(x)}{\partial x}\frac{\varepsilon_1^2}{\sigma_1} \tag{2.38}$$

Compare this result to (2.9). Note that there is nothing in (2.9) that corresponds to the last term in the above expression. Also recall we are evaluating this quantity in the limit $\sigma_1 \to 0$. In this case the last term, above, is not well defined. We refer to (2.33) and (2.11) and write $\varepsilon_1^2/\sigma_1$ to obtain,

$$\frac{\varepsilon_1^2}{\sigma_1} = \frac{\varepsilon_1^2}{\left[\varepsilon_1^2 - \varepsilon_0^2 + 2i\varepsilon_0\tau + \tau^2\right]} \tag{2.39}$$

This quantity is dependent on how $\varepsilon_0$, $\varepsilon_1$, and $\tau$ go to zero. For example if $\tau$ and $\varepsilon_0$ go to zero much faster than $\varepsilon_1$ then the above quantity will be $\varepsilon_1^2/\sigma_1 \to \varepsilon_1^2/\varepsilon_1^2 = 1$. However if $\varepsilon_1$ approaches zero faster than $\tau$ and $\varepsilon_0$ then $\varepsilon_1^2/\sigma_1 \to 0$. If we set $\tau = 0$ and $\varepsilon_0 = \varepsilon_1 \to 0$ then $\varepsilon_1^2/\sigma_1 \to \infty$. Therefore in order to properly evaluate the above expressions we must know how these quantities approach zero.

### 3. Kinetic energy density for a static potential.

In this section we will provide another example where a calculation using the method of point split regularization results in an expression with is dependent on how the quantities $\varepsilon$ and $\tau$ approach zero.

We will calculate the kinetic energy density for a zero mass scalar field in 1-1 dimensional space-time where the scalar potential $\lambda V(x)$ is given by,

$$\lambda V(x) = \begin{cases} \lambda & \text{for } |x| < a \\ 0 & \text{for } |x| > a \end{cases} \tag{3.1}$$

The field operator satisfies (2.1) with $\lambda V(x)$ given by the above expression and $m = 0$. For this case the field operator is given by,

$$\hat{\varphi}_\lambda(x,t) = \sum_{j=1,2} \int_0^\infty \left( \hat{a}_{\lambda,j\omega} f_{\lambda,j\omega}(x,t) + \hat{a}^*_{\lambda,j\omega} f^*_{\lambda,j\omega}(x,t) \right) d\omega \tag{3.2}$$

where $\hat{a}_{\lambda,j\omega}$ and $\hat{a}^*_{\lambda,j\omega}$ are the destruction and creation operators, respectively. They obey the commutation relationships $\left[ \hat{a}_{\lambda,j\omega}, \hat{a}^*_{\lambda,j'\omega'} \right] = \delta(\omega - \omega') \delta_{j'j}$ with all other commutations being zero. As was discussed in Ref [10] the $f_{\lambda,j\omega}(x,t)$ are given by,

$$f_{\lambda,j\omega}(x,t) = \frac{e^{-i\omega t}}{\sqrt{2\pi\omega}} \chi_{j\omega}(x) \tag{3.3}$$

where the function $\chi_{j\omega}(x)$ is the eigensolution to the equation,

$$-\omega^2 \chi_{j\omega} - \frac{\partial^2 \chi_{j\omega}}{\partial x^2} + \lambda V(x) \chi_{j\omega} = 0 \tag{3.4}$$

In the above expression $\omega$ takes on all values from $0$ to $\infty$ and $j = 1, 2$ where $j = 1$ stands for the symmetric solutions and $j = 2$ stands for the anti-symmetric solutions. For the free field case, for which $\lambda = 0$, we have that,

$$f_{0,j\omega}(x,t) = \frac{e^{-i\omega t}}{\sqrt{2\pi\omega}} \begin{cases} \cos(\omega x) & \text{for } j = 1 \\ \sin(\omega x) & \text{for } j = 2 \end{cases} \tag{3.5}$$

For the massless case the two point function will have an infrared divergence so we will not use it. Instead we will evaluate the point split kinetic energy density using (2.12) and (2.14). First define the quantity,

$$\xi_{\lambda,j\omega}(y;y') = \frac{1}{4} \left[ \left( \frac{\partial f_{\lambda,j\omega}(y)}{\partial y_0} \frac{\partial f^*_{\lambda,j\omega}(y')}{\partial y'_0} + \frac{\partial f_{\lambda,j\omega}(y)}{\partial y_1} \frac{\partial f^*_{\lambda,j\omega}(y')}{\partial y'_1} \right) + c.c. \right] \tag{3.6}$$

This can be considered to be the point split kinetic energy density of the $j\omega - th$ mode. Next define,

$$\xi_{\lambda,\omega}(y;y') = \sum_{j=1,2} \xi_{\lambda,j\omega}(y;y') \tag{3.7}$$

The total kinetic energy density is then given by,

$$T_{00,\lambda}(x,t;\varepsilon,\tau) = \int_0^\infty \xi_{\lambda,\omega}(y,y') e^{-\omega \tau} d\omega \tag{3.8}$$

Note that we have included a factor $e^{-\omega\tau}$ where $\tau \to 0$. This factor serves as a frequency cutoff and corresponds to the term $\tau$ in the definition of $\sigma_1$ in Eq. (2.33). For the free field case the kinetic energy density is obtained by setting $\lambda = 0$ in (3.8) to obtain,

$$T_{00,0}(x,t;\varepsilon,\tau) = \int_0^\infty \xi_{0,\omega}(y,y') e^{-\omega\tau} d\omega \tag{3.9}$$

The renormalized kinetic energy density is then,

$$T_{00R}(x,t;\varepsilon,\tau) = T_{00,\lambda}(x,t;\varepsilon,\tau) - T_{00,0}(x,t;\varepsilon,\tau) \tag{3.10}$$

Use (3.8) and (3.9) in the above to obtain,

$$T_{00,R}(y;y') = \int_0^\infty \left( \xi_{\lambda,\omega}(y;y') - \xi_{0,\omega}(y;y') \right) e^{-\omega\tau} d\omega \tag{3.11}$$

In order to evaluate this we need solve for the eigensolutions $\chi_{\lambda,j\omega}(x)$. This is done in the Appendix (Also see Ref [10]). We will evaluate the kinetic energy density in the region $|x| < a$. This is the region where the scalar potential equals $\lambda$. From the Appendix, for this region, we evaluate $\xi_{\lambda,\omega}(y;y')$ and $\xi_{0,\omega}(y;y')$ as,

$$\xi_{\lambda,\omega}(y;y') = \left[ \left( \frac{e^{-i\omega(y_0-y_0')} + c.c}{8\pi} \right) \left( \begin{array}{c} \left( A_{1\omega}^2 + A_{2\omega}^2 \right) \left( \omega - \frac{\lambda}{2\omega} \right) \cos\left( (y_1 - y_1') \sqrt{\omega^2 - \lambda} \right) \\ + \left( A_{2\omega}^2 - A_{1\omega}^2 \right) \frac{\lambda}{2\omega} \cos\left( (y_1 + y_1') \sqrt{\omega^2 - \lambda} \right) \end{array} \right) \right] \tag{3.12}$$

and,

$$\xi_{0,\omega}(y;y') = \left( \frac{e^{i\omega(y_0-y_0')} + c.c.}{4\pi} \right) \omega \cos\left[ \omega(y_1 - y_1') \right] \tag{3.13}$$

where $A_{1\omega}$ and $A_{2\omega}$ are given in the Appendix and $c.c.$ means to take the complex conjugate of the proceeding term. Define the quantities,

$$R_\omega(y;y') = \frac{\lambda}{4\pi}(y_1 - y_1') \cos\left[ \omega(y_0 - y_0') \right] \sin\left[ \omega(y_1 - y_1') \right] \tag{3.14}$$

and,

$$S_\omega(y;y') = \left( \xi_{\lambda,\omega}(y;y') - \xi_{0,\omega}(y;y') \right) - R_\omega(y;y') \tag{3.15}$$

Use these results in (3.11) to obtain,

$$T_{00R}(x,t;\varepsilon,\tau) = \int_0^\infty S_\omega(y;y')e^{-\omega\tau}d\omega + \int_0^\infty R_\omega(y;y')e^{-\omega\tau}d\omega \qquad (3.16)$$

The quantity $S_\omega(y;y')$ falls off sufficiently fast so that in the limit that $\varepsilon = (\varepsilon_0,\varepsilon_1) \to 0$ and $\tau \to 0$ the first integral on the right in the above expression will be independent of $\varepsilon$ and $\tau$. Therefore any possible dependency on $\varepsilon_0$, $\varepsilon_1$, and $\tau$ as these terms approach zero is from the second integral. The second integral is readily evaluated as,

$$\int_0^\infty R_\omega(y;y')e^{-\omega\tau}d\omega = \frac{\lambda}{8\pi}\left[\left(\frac{\varepsilon_1^2}{\varepsilon_1^2 - \varepsilon_0^2 - 2i\tau\varepsilon_0 + \tau^2}\right) + \text{c.c.}\right] \qquad (3.17)$$

where this quantity is calculate in the limit that $(\varepsilon_0,\varepsilon_1,\tau) \to 0$.

Note that we are in the same situation as in Section 2. This expression depends on the way in which the quantities $(\varepsilon_0,\varepsilon_1,\tau)$ approach zero. Therefore, unless we know how to take these expressions to zero, we cannot evaluate the point split kinetic energy density.

**4. Field operator for a time dependent potential..**

In order to resolve this dilemma we will consider one more example. Consider a massive scalar field in the presence of a time dependent scalar potential $V(t)$. In case this the mode solutions satisfy,

$$\frac{\partial^2 f_k(x,t)}{\partial t^2} - \frac{\partial^2 f_k(x,t)}{\partial x^2} + (m^2 + V(t))f_k(x,t) = 0 \qquad (4.1)$$

For this problem we will assume periodic boundary conditions of period $L$ so that, $f_k(x,t) = f_k(x+L,t)$. These solutions are given by,

$$f_k(x,t) = \frac{s_k(t)e^{-ikx}}{\sqrt{2\omega L}} \qquad (4.2)$$

where $\omega = \sqrt{k^2 + m^2}$ and $s_k(t)$ satisfies,

$$\frac{\partial^2 s_k(t)}{\partial t^2} + (\omega^2 + V(t))s_k(t) = 0 \qquad (4.3)$$

In order to satisfy the periodic boundary conditions we have,

$$k = \frac{2\pi n}{L} \qquad (4.4)$$

where $n$ is an integer. Let the potential be given by

$$V(t) = \lambda \theta(t) \tag{4.5}$$

where $\lambda$ is a constant. The solutions to the above is given by,

$$s_k(t) = \begin{cases} e^{-i\omega t} & ; t < 0 \\ A_k e^{iEt} + B_k e^{-iEt} & ; t > 0 \end{cases} \tag{4.6}$$

where $E = \sqrt{\omega^2 + \lambda} = \sqrt{k^2 + m^2 + \lambda}$. The constants $A_k$ and $B_k$ are derived from the boundary conditions at $t = 0$. These are that $s_k(t)$ and $ds_k(t)/dt$ are continuous at $t = 0$. From these conditions we obtain,

$$A_k = \frac{1}{2}\left(1 - \frac{\omega}{E}\right) \text{ and } B_k = \frac{1}{2}\left(1 + \frac{\omega}{E}\right) \tag{4.7}$$

**5. Mode Regularization.**

In this section we will determine the kinetic energy density for the system defined in the previous section using "mode" regularization. Mode regulation is based on the concept that we can trace the evolution through time of each individual mode $f_k(x,t)$ which will allow us to calculate the change in the kinetic energy density of each mode. The total kinetic energy density is obtained by adding up all these changes.

The kinetic energy density of the *k-th* mode is,

$$\xi_{M\lambda,k}(x,t) = \frac{1}{2}\left(\left|\frac{\partial f_k(x,t)}{\partial t}\right|^2 + \left|\frac{\partial f_k(x,t)}{\partial x}\right|^2 + m^2 |f_k(x,t)|^2\right) \tag{5.1}$$

At $t < 0$, when $\lambda = 0$ before the application of the scalar potential, the kinetic energy density of the *k-th* mode is,

$$\xi_{M0,k}(x) = \frac{\omega}{2L} \tag{5.2}$$

Therefore the change in the kinetic energy density of this mode due to the application of the scalar potential is,

$$\Delta \xi_{M,k}(x,t) = \xi_{M\lambda,k}(x,t) - \frac{\omega}{2L} \tag{5.3}$$

Assume $t > 0$ and use (4.2) and (4.6) in (5.1) to obtain,

$$\xi_{M\lambda,k}(x,t) = \frac{1}{8\omega L}\left\{\left[\left[1+\left(\frac{\omega}{E}\right)^2\right](\lambda+2\omega^2)\right] - \lambda\left[1-\left(\frac{\omega}{E}\right)^2\right]\cos(2Et)\right\} \quad (5.4)$$

The change in the kinetic energy density of the *k-th* mode is then,

$$\Delta\xi_{M,k}(x,t) = \frac{1}{8\omega L}\left(\frac{\lambda^2}{E^2}\right)(1-\cos(2Et)) \quad (5.5)$$

According to mode regularization the total renormalized kinetic energy density is simply the sum of the change in the kinetic energy density of each mode to obtain,

$$T_{M00R}(x,t) = \sum_k \Delta\xi_{M,k}(x,t) \quad (5.6)$$

(Note the subscript "M" in $T_{M00R}$ denotes that this quantity is calculated using mode regularization). In the limit $L \to \infty$ we can write $\sum_n \to \int \frac{L}{2\pi}dk$ to obtain,

$$T_{M00R}(x,t) = \frac{1}{16\pi}\int_{-\infty}^{+\infty}\frac{1}{\omega}\left(\frac{\lambda^2}{E^2}\right)(1-\cos(2Et))dk \quad (5.7)$$

This expression converges due to the fact that the integrand falls off sufficiently fast as $|k| \to \infty$. Therefore, using mode regularization, we can obtain a well defined and finite expression for the kinetic energy densisty.

## 6. Point Split Regularization.

In this section we will calculate the kinetic energy density using point split regularization and compare the results. For this problem the two point function is given by,

$$G(y;y';\tau) = \sum_k g_k(y;y')e^{-\omega\tau} \quad (6.1)$$

where,

$$g_{\lambda,k}(y;y') = \frac{1}{4\omega L}\left[s_{\lambda,k}(y_0)s^*_{\lambda,k}(y'_0)e^{ik(y_1-y'_1)} + c.c.\right] \quad (6.2)$$

As in the prior discussions a frequency cutoff factor is used and it is assumed that $\tau \to 0$. The point split kinetic energy density is then,

$$T_{00,\lambda}(x,t;\varepsilon,\tau) = \frac{1}{2}\left(\frac{\partial}{\partial y_0}\frac{\partial}{\partial y'_0} + \frac{\partial}{\partial y_1}\frac{\partial}{\partial y'_1} + m^2\right)G(y;y';\tau) \quad (6.3)$$

Using (6.1) and (6.2) this can be written as,

$$T_{00,\lambda}(x,t;\varepsilon,\tau) = \sum_k \xi_{\lambda,k}(y;y') e^{-\omega\tau} \tag{6.4}$$

where,

$$\xi_{\lambda,k}(y;y') = \frac{1}{2}\left(\frac{\partial}{\partial y_0}\frac{\partial}{\partial y'_0} + \frac{\partial}{\partial y_1}\frac{\partial}{\partial y'_1} + m^2\right) g_{\lambda,k}(y;y') \tag{6.5}$$

The renormalized point split kinetic energy density is,

$$T_{00R}(x,t;\varepsilon,\tau) = T_{00,\lambda}(x,t;\varepsilon,\tau) - T_{00,0}(x,t;\varepsilon,\tau) \tag{6.6}$$

where,

$$T_{00,0}(x,t;\varepsilon,\tau) = \sum_k \xi_{0,k}(y;y') e^{-\omega\tau} \tag{6.7}$$

Use (6.2) and (4.6) in (6.5) to obtain, for $t > 0$,

$$\xi_{\lambda,k}(y;y') = \frac{1}{8\omega L}\left[\begin{pmatrix}(2\omega^2+\lambda)\left(A_k^2 e^{iE(y_0-y'_0)} + B_k^2 e^{-iE(y_0-y'_0)}\right) \\ -2\lambda A_k B_k \cos[E(y_0+y'_0)]\end{pmatrix} e^{ik(y_1-y'_1)} + c.c.\right] \tag{6.8}$$

$\xi_{0,k}(y;y')$ is obtained by setting $\lambda = 0$ in the above to yield,

$$\xi_{0,k}(y;y') = \frac{\omega}{4L}\left[e^{-i\omega(y_0-y'_0)} e^{ik(y_1-y'_1)} + c.c.\right] \tag{6.9}$$

Define,

$$R_k(y;y') = \frac{i\lambda(y_0-y'_0)}{8L}\left[e^{-i\omega(y_0-y'_0)} e^{ik(y_1-y'_1)} - c.c.\right] \tag{6.10}$$

and,

$$S_k(y;y') = \left(\xi_{\lambda,k}(y;y') - \xi_{0,k}(y;y')\right) - R_k(y;y') \tag{6.11}$$

Use the above results in (6.6) to obtain and let $L \to \infty$ to obtain,

$$T_{00R}(x,t;\varepsilon,\tau) = \frac{1}{2\pi}\int_{-\infty}^{+\infty} S_k(y;y') e^{-\omega\tau} dk + \frac{1}{2\pi}\int_{-\infty}^{+\infty} R_k(y;y') e^{-\omega\tau} dk \tag{6.12}$$

Let us consider the first integral on the right. For large $|k|$ the quantity $S_k(y;y')$ falls off sufficiently fast so that in the limit that $\varepsilon, \tau \to 0$ we can show that,

$$T_{M00R}(x,t) \underset{\varepsilon,\tau\to 0}{=} \frac{1}{2\pi}\int_{-\infty}^{+\infty} S_k(y;y') e^{-\omega\tau} dk \tag{6.13}$$

where $T_{M00R}(x,t)$ is the kinetic energy density as determined by mode regularization in the last section (see Eq. (5.7). Using these results we obtain,

$$T_{00R}(x,t;\varepsilon,\tau) = T_{M00R}(x,t) + D(\varepsilon_0,\varepsilon_1,\tau) \tag{6.14}$$

where,

$$D(\varepsilon_0,\varepsilon_1,\tau) = \frac{1}{2\pi}\int_{-\infty}^{+\infty} R_k(y;y')e^{-\omega\tau}dk \tag{6.15}$$

In the limit that $(\varepsilon_0,\varepsilon_1) \to 0$ we can replace $\omega$ with $|k|$ to obtain,

$$D(\varepsilon_0,\varepsilon_1,\tau) = -\frac{\lambda\varepsilon_0}{8\pi}\left(\frac{\varepsilon_0-i\tau}{\left[(\varepsilon_1^2-\varepsilon_0^2)+2i\varepsilon_0\tau+\tau^2\right]} + \frac{\varepsilon_0+i\tau}{\left[(\varepsilon_1^2-\varepsilon_0^2)-2i\varepsilon_0\tau+\tau^2\right]}\right) \tag{6.16}$$

As in the previous example this quantity is dependent on how $\varepsilon$ and $\tau$ approach zero. If this term was zero then mode regularization and point split regularization would give the same result. In all our examples the terms of this type will be eliminated if we assume that $\varepsilon_0$ and $\varepsilon_1$ go to zero much faster than $\tau$. In this case we can set $\varepsilon_0$ and $\varepsilon_1$ equal to zero in (6.16) to obtain $D(\varepsilon_0,\varepsilon_1,\tau) = 0$.

## 7. Flanagan's proof of the spatial quantum inequality.

In this section we will examine certain elements of Flanagan's [5] proof of the spatial quantum inequality and show how they are impacted by results of the previous discussion. Consider a zero mass scalar field in 1-1D space-time with the scalar potential equal to zero. The field operator can be decomposed as,

$$\hat{\varphi}(x,t) = \hat{\varphi}_R(v) + \hat{\varphi}_L(u) \tag{7.1}$$

where $v = x-t$ and $u = x+t$ and where $\hat{\varphi}_R(v)$ acts on the right moving section and $\hat{\varphi}_L(v)$ acts on the left moving sector. In the following discussion we will consider the right moving sector only. For this sector we define the point split kinetic energy operator as,

$$\hat{T}_{vv}\left(v,\bar{v};[\hat{\varphi}_R(v)]\right) = \frac{d\hat{\varphi}_R(v)}{dv}\cdot\frac{d\hat{\varphi}_R(\bar{v})}{d\bar{v}} \tag{7.2}$$

For the "free field" the state vector $\hat{\varphi}_R(v)$ is given by,

$$\hat{\varphi}_R(v) = \frac{1}{\sqrt{2\pi}} \int_0^\infty d\omega \frac{1}{\sqrt{2\omega}} \left( \hat{a}_\omega e^{-i\omega v} + \hat{a}_\omega^* e^{+i\omega v} \right) \tag{7.3}$$

The vacuum state associated with this state vector is $|0\rangle$ where $\hat{a}_\omega |0\rangle = 0$. The point split energy density of the vacuum state $|0\rangle$ is then,

$$\langle 0 | \hat{T}_{vv} (v, \bar{v}; [\hat{\varphi}_R(v)]) | 0 \rangle = \langle 0 | \frac{d\hat{\varphi}_R(v)}{dv} \cdot \frac{d\hat{\varphi}_R(\bar{v})}{d\bar{v}} | 0 \rangle \tag{7.4}$$

Use (7.2) and (7.3) in the above and add the frequency cutoff factor to obtain,

$$\langle 0 | \hat{T}_{vv} (v, \bar{v}; [\hat{\varphi}_R(v)]) | 0 \rangle = \frac{1}{4\pi} \int_0^\infty \omega e^{-i\omega(v-\bar{v})} e^{-\omega\tau} d\omega \tag{7.5}$$

This can be readily integrated to obtain,

$$\langle 0 | \hat{T}_{vv} (v, \bar{v}; [\hat{\varphi}_R(v)]) | 0 \rangle = -\frac{1}{4\pi \left( (v-\bar{v}) - i\tau \right)^2} \tag{7.6}$$

This can also be written as,

$$\langle 0 | \hat{T}_{vv} (v, \bar{v}; [\hat{\varphi}_R(v)]) | 0 \rangle = -\frac{1}{4\pi} \frac{\partial}{\partial \bar{v}} \frac{\partial}{\partial v} \left[ \ln \left( (v - \bar{v}) - i\tau \right) \right] \tag{7.7}$$

Following Flanagan we assume the existence of a unitary operator $\hat{S}$ which acts on the state vector $\hat{\varphi}_R(v)$ to obtain,

$$\hat{S}^\dagger \hat{\varphi}_R(v) \hat{S} = \hat{\varphi}_R(V(v)) \tag{7.8}$$

In this case,

$$\langle 0 | \hat{S}^\dagger \hat{T}_{vv} (v, \bar{v}; [\hat{\varphi}_R(v)]) \hat{S} | 0 \rangle = \langle 0 | \hat{T}_{vv} (v, \bar{v}; [\hat{\varphi}_R(V(v))]) | 0 \rangle \tag{7.9}$$

and,

$$\langle 0 | \hat{T}_{vv} (v, \bar{v}; [\hat{\varphi}_R(V(v))]) | 0 \rangle = -\frac{1}{4\pi} \frac{\partial}{\partial \bar{v}} \frac{\partial}{\partial v} \left[ \ln \left( [V(v) - V(\bar{v})] - i\tau \right) \right] \tag{7.10}$$

Define,

$$\Delta(v, \bar{v}; \tau) = \langle 0 | \hat{T}_{vv} (v, \bar{v}; [\hat{\varphi}_R(v)]) | 0 \rangle - \langle 0 | \hat{T}_{vv} (v, \bar{v}; [\hat{\varphi}_R(V(v))]) | 0 \rangle \tag{7.11}$$

The quantity $\Delta(v, \bar{v}; \tau)$ is defined this way in order to correspond to the quantity $\Delta(v)$ which is defined in [5].

Use (7.7) and (7.10) in the above to obtain,

$$\Delta(v,\bar{v};\tau) = \frac{1}{4\pi}\frac{\partial}{\partial \bar{v}}\frac{\partial}{\partial v}\left[\ln\left(\left[V(v)-V(\bar{v})\right]-i\tau\right) - \ln\left(\left[v-\bar{v}\right]-i\tau\right)\right] \quad (7.12)$$

This becomes,

$$\Delta(v,\bar{v};\tau) = \frac{1}{4\pi}\left\{\frac{\left(dV(v)/dv\right)\left(dV(\bar{v})/d\bar{v}\right)}{\left(\left[V(v)-V(\bar{v})\right]-i\tau\right)^2} - \frac{1}{\left(\left[v-\bar{v}\right]-i\tau\right)^2}\right\} \quad (7.13)$$

Next we want to evaluate this in the limit that $\bar{v} \to v$ and $\tau \to 0$. In Flanagan's formulation $\tau$ does not appear so we set $\tau = 0$. In this case (7.13) becomes,

$$\Delta(v,\bar{v}) = \frac{1}{4\pi}\left\{\frac{\left(dV(v)/dv\right)\left(dV(\bar{v})/d\bar{v}\right)}{\left(V(v)-V(\bar{v})\right)^2} - \frac{1}{\left(v-\bar{v}\right)^2}\right\} \quad (7.14)$$

In the limit $\bar{v} \to v$ the quantity $V(\bar{v})$ can be expanded in a Taylor series as,

$$V(\bar{v}) = V(v) + \frac{dV(v)}{dv}(\bar{v}-v) + \frac{1}{2!}\frac{d^2V(v)}{dv^2}(\bar{v}-v)^2 + \frac{1}{3!}\frac{d^3V(v)}{dv^3}(\bar{v}-v)^3 + \ldots \quad (7.15)$$

Following Flanagan [5] define,

$$\Delta(v) \underset{\bar{v}\to v}{=} \Delta(v,\bar{v}) \quad (7.16)$$

Use (7.15) in (7.14) to obtain,

$$\Delta(v) = \frac{1}{4\pi}\left[\frac{V'''(v)}{6V'(v)} - \frac{V''(v)^2}{4V'(v)^2}\right] \quad (7.17)$$

This is identical to Flanagan's result (See Eq. 2.17 of [5]).

However consider the situation where $\bar{v} \to v$ much faster than $\tau \to 0$. For this case define,

$$\Delta(v;\tau) \underset{\bar{v}\to v}{=} \Delta(v,\bar{v};\tau) \quad (7.18)$$

Therefore, since $|\tau| \ll |v-\bar{v}|$ in the limit $\bar{v} \to v$ we can set $\bar{v} = v$ in the denominators in (7.13) and (7.13) can be evaluated as,

$$\Delta(v;\tau) = -\frac{1}{4\pi\tau^2}\left\{\left(\frac{dV(v)}{dv}\right)^2 - 1\right\} \quad (7.19)$$

Note that this is significantly different from $\Delta(v)$. If $\Delta(v;\tau)$ is used instead of $\Delta(v)$ then Flanagan's proof of the quantum inequalities will fail.

## 8. Conclusion.

We have worked a number of examples and shown that when point splitting is used to calculate the kinetic energy density the final result includes a term that is dependent on how the quantities $\varepsilon_0$, $\varepsilon_1$, and $\tau$ go to zero. We have proposed that $\varepsilon_0$ and $\varepsilon_1$ should go to zero much faster than $\tau$. This has implications regarding Flanagan's proof of the quantum interest conjecture. The proof, as presented in [5], does not include $\tau$ therefore $\tau$ has implicitly been set equal to zero. However if $\tau$ in included and the quantity $\bar{v} - v \to 0$ much faster than $\tau \to 0$ then the proof fails.

## Appendix.

We will determine the mode solutions $\chi_{j\omega}(x)$. We start by evaluating the symmetric solutions, with $j = 1$. For the region where $|x| < a$, which will be specified as region $I$, we have,

$$\chi_{1\omega}(x; I) = A_{1\omega} \cos\left(x\sqrt{\omega^2 - \lambda}\right) \tag{8.1}$$

For $|x| > a$, which is specified as region $II$, we have,

$$\chi_{1\omega}(x; II) = \cos\left(\omega|x| + \delta_{1\omega}\right) \tag{8.2}$$

For the anti-symmetric solutions, where $j = 2$, we have the following. For region I where $|x| < a$ we have,

$$\chi_{2\omega}(x; I) = A_{2\omega} \sin\left(x\sqrt{\omega^2 - \lambda}\right) \tag{8.3}$$

For region $II$ where $|x| > a$ we have,

$$\chi_{2\omega}(x; II) = \sin\left(\omega x + \delta_{2\omega}\varepsilon(x)\right) \tag{8.4}$$

where $\varepsilon(x) = +1$ for $x > 0$ and $\varepsilon(x) = -1$ for $x < 0$.

The boundary conditions at $x = \pm a$ are that $\chi_{j\omega}(x)$ and its first derivative $d\chi_{j\omega}(x)/dx$ are continuous across the boundary. For the symmetric solutions this yields,

$$\cos(\omega a + \delta_{1\omega}) = A_{1\omega} \cos\left(a\sqrt{\omega^2 - \lambda}\right) \tag{8.5}$$

and,

$$-\omega \sin(\omega a + \delta_{1\omega}) = -A_{1\omega} \sqrt{\omega^2 - \lambda} \sin\left(a\sqrt{\omega^2 - \lambda}\right) \tag{8.6}$$

From these equations we obtain,

$$A_{1\omega}^2 = \frac{1}{\left[1 - \frac{\lambda}{\omega^2} \sin^2\left(a\sqrt{(\omega^2 - \lambda)}\right)\right]} \tag{8.7}$$

For the anti-symmetric solutions we obtain,

$$\sin(\omega a + \delta_{1\omega}) = A_{2\omega} \sin\left(a\sqrt{\omega^2 - \lambda}\right) \tag{8.8}$$

and

$$\omega \cos(\omega a + \delta_{1\omega}) = A_{2\omega} \sqrt{\omega^2 - \lambda} \cos\left(a\sqrt{\omega^2 - \lambda}\right) \tag{8.9}$$

and,

$$A_{2\omega}^2 = \frac{1}{\left[1 - \frac{\lambda}{\omega^2} \cos^2\left(a\sqrt{(\omega^2 - \lambda)}\right)\right]} \tag{8.10}$$

The kinetic energy density $\xi_{\lambda, j\omega}(y; y')$ of a given mode. This will be determined in the region $I$ where $|x| < a$. First use (3.3) in (3.6) to obtain,

$$\xi_{\lambda, j\omega}(y; y') = \left[\left(\frac{e^{i\omega(y_0 - y_0')}}{4\pi\omega}\right)\left(\omega^2 \chi_{j\omega}(y_1) \chi_{j\omega}^*(y_1') + \frac{\partial \chi_{j\omega}(y_1)}{\partial y_1} \frac{\partial \chi_{j\omega}^*(y_1')}{\partial y_1'}\right)\right] + c.c \tag{8.11}$$

Next use these results to obtain,

$$\xi_{\lambda, 1\omega}(y; y') = \left[A_{1\omega}^2 \left(\frac{e^{i\omega(y_0 - y_0')}}{4\pi\omega} + c.c\right)\left(\begin{array}{c}\omega^2 \cos\left((y_1 - y_1')\sqrt{\omega^2 - \lambda}\right) \\ -\lambda \sin\left(y_1 \sqrt{\omega^2 - \lambda}\right) \sin\left(y_1' \sqrt{\omega^2 - \lambda}\right)\end{array}\right)\right] \tag{8.12}$$

and,

$$\xi_{\lambda,2\omega}(y;y') = \left[ A_{2\omega}^2 \left( \frac{e^{i\omega(y_0-y_0')}+c.c}{4\pi\omega} \right) \begin{pmatrix} \omega^2 \cos\left((y_1-y_1')\sqrt{\omega^2-\lambda}\right) \\ -\lambda \cos\left(y_1\sqrt{\omega^2-\lambda}\right)\cos\left(y_1'\sqrt{\omega^2-\lambda}\right) \end{pmatrix} \right] \quad (8.13)$$

Use these in (3.7) to obtain (3.12) in the text.

**References.**

13. Y. Decanini and A. Folacci. "Hadamard renormalization of the stress-energy tensor for a quantized scalar field in a general spacetime of arbitrary dimension", Phys. Rev. D78:044025,2008. arXiv:gr-qc/0512118.
14. R. M. Wald, "Quantum Field Theory in Curved Spacetime and Black Hole Thermodynamics" University of Chicago Press, Chicago (1994).
15. N.D. Birrell and P.C.W Davies, "Quantum fields in curved space". Cambridge University Press, Cambridge (1982).
21